\documentclass[letterpaper]{article} 
\usepackage{aaai2026}  
\usepackage{times}  
\usepackage{helvet}  
\usepackage{courier}  
\usepackage[hyphens]{url}  
\usepackage{graphicx} 
\usepackage{natbib}  
\usepackage{caption} 
\usepackage{algorithm}
\usepackage{algorithmic}
\usepackage{subfigure}
\usepackage{multirow}
\usepackage{amsmath}
\usepackage{amssymb}
\usepackage{booktabs}

\usepackage{newfloat}
\usepackage{listings}
\DeclareCaptionStyle{ruled}{labelfont=normalfont,labelsep=colon,strut=off} 
\floatstyle{ruled}
\newfloat{listing}{tb}{lst}{}
\floatname{listing}{Listing}
\title{Wavelet Enhanced Adaptive Frequency Filter for \\
Sequential Recommendation}
\author{
    Huayang Xu\textsuperscript{\rm 1}\equalcontrib,
    Huanhuan Yuan\textsuperscript{\rm 1,2}\equalcontrib,
    Guanfeng Liu\textsuperscript{\rm 2},
    Junhua Fang\textsuperscript{\rm 1},
    Lei Zhao\textsuperscript{\rm 1},
    Pengpeng Zhao\textsuperscript{\rm 1}\thanks{Corresponding author.}
}
\affiliations{
    \textsuperscript{\rm 1}School of Computer Science and Technology, Soochow University, China\\
    \textsuperscript{\rm 2}School of Computing, Macquarie University, Australia\\

    hyxu2001@stu.suda.edu.cn, hhyuan@stu.suda.edu.cn, guanfeng.liu@mq.edu.au, \{jhfang, zhaol, ppzhao\}@suda.edu.cn

}

\usepackage{bibentry}

\begin{document}

\maketitle

\begin{abstract}
Sequential recommendation has garnered significant attention for its ability to capture dynamic preferences by mining users' historical interaction data. Given that users' complex and intertwined periodic preferences are difficult to disentangle in the time domain, recent research is exploring frequency domain analysis to identify these hidden patterns. However, current frequency-domain-based methods suffer from two key limitations: (i) They primarily employ static filters with fixed characteristics, overlooking the personalized nature of behavioral patterns; (ii) While the global discrete Fourier transform excels at modeling long-range dependencies, it can blur non-stationary signals and short-term fluctuations. To overcome these limitations, we propose a novel method called \textbf{W}avelet \textbf{E}nhanced  \textbf{A}daptive Frequency Filter for Sequential \textbf{Rec}ommendation (WEARec).  Specifically, it consists of two vital modules: dynamic frequency-domain filtering and wavelet feature enhancement. The former is used to dynamically adjust filtering operations based on behavioral sequences to extract personalized global information, and the latter integrates wavelet transform to reconstruct sequences, enhancing blurred non-stationary signals and short-term fluctuations. Finally, these two modules work synergistically to achieve comprehensive performance and efficiency optimization in long sequential recommendation scenarios. Extensive experiments on four widely-used benchmark datasets demonstrate the superiority of WEARec.
\end{abstract}


\begin{links}
\link{Code}{https://github.com/xhy963319431/WEARec}
\end{links}

\section{Introduction}
Sequential Recommendation (SR) plays a crucial role in e-commerce applications by capturing users' dynamic interest shifts through their historical interaction data \cite{music,music2}. The remarkable success of the transformer architecture in Natural Language Processing (NLP) \cite{attention} and Computer Vision (CV) \cite{vit} has led to significant advancements in sequential recommendation \cite{sasrec,feature,contrastive}. This has directly inspired a multitude of sequential recommendation models based on self-attention \cite{bert,duo,s3}.  However, items in user interactions are typically chronologically entangled and inherently noisy \cite{slime,fea}. Consequently, it is challenging for models to directly discern changes in behavioral preferences from raw sequences within the temporal domain \cite{fmlp}. 
\begin{figure}[tbp]

\centering
\centerline{\includegraphics[width=0.48\textwidth]{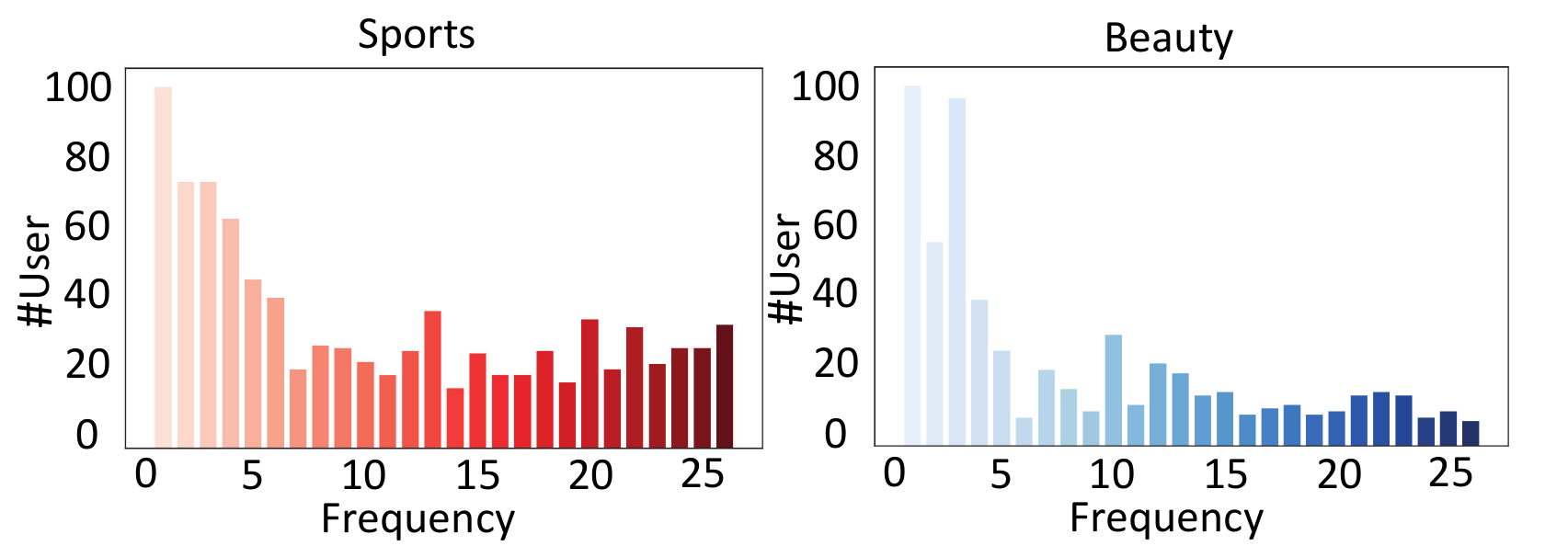}}
\caption{Number of users uniquely driven by each frequency component in the Sports and Beauty datasets.}
\label{Fig.1}
\end{figure}

To address this limitation, recent research has begun exploring frequency-domain approaches to replace self-attention mechanisms \cite{gfnet,fft,glfnet}. By decomposing user sequences into different frequency components ($e.g.$, high-frequency and low-frequency signals) using Fourier transform, periodic patterns that are difficult to identify in the time domain can be effectively captured \cite{fmlp,slime,bsa,famous}. For example, FMLPRec \cite{fmlp} pioneered frequency domain processing of sequential data, replacing the self-attention mechanism with learnable filters. SLIME4Rec \cite{slime} proposed a frequency ramp structure, which considers different frequency bands for each layer. BSARec \cite{bsa} used a frequency domain re-tuning component as an inductive bias for self-attention.

However, despite their success in SR, existing frequency-domain sequential recommendation models have two key limitations. First, existing methods typically apply a static, fixed-pattern filter to all frequency components, which uniformly processes all user sequences, ignoring the personalized nature of behavioral patterns \cite{bsa,famous}. 
This unified filtering approach is susceptible to the influence of dominant users in the dataset and fails to consider the characteristics of individual users. 
In fact, users often exhibit diverse behavioral patterns: some user behaviors follow long-term preferences ($e.g.$, low-frequency signals), while other user behaviors show the opposite trend \cite{slime,famous}. To illustrate this, we trained a classic sequential recommendation model ($i.e.$, FMLPRec \cite{fmlp}) on the Beauty and Sports datasets, and replaced its learnable filters with specific band-pass filters. By statistically analyzing how many users could be correctly predicted by specific frequency components, we could identify which users were driven by particular frequency components. The results, shown in Figure \ref{Fig.1}, indicate that users exhibit diverse behavioral patterns, with each focusing on different frequencies. This emphasizes the importance of developing personalized filtering models to capture individualized user behavioral patterns.

The second limitation is related to the low-pass filtering characteristic of frequency-domain filters. The Discrete Fourier Transform (DFT) analyzes signal components globally, primarily serving as a global, rather than local, frequency extraction method. While the global DFT excels at capturing long-range dependencies in current frequency-domain recommendation models \cite{fmlp,slime}, it struggles to capture the local temporal features of high-frequency interactions and short-term points of interest \cite{wave}. For instance, FMLPRec has been shown to essentially act as a low-pass filter \cite{bsa}. Although SLIME4Rec attempts to balance high-frequency and low-frequency representations through its hierarchical learning mechanism, the model still tends to learn low-frequency components within the hierarchical frequency bands.

To overcome these challenges, we propose \textbf{W}avelet \textbf{E}nhanced \textbf{A}daptive Frequency Filter for Sequential \textbf{Rec}om-
mendation (WEARec). WEARec consists of two key modules: Dynamic {F}requency {F}iltering (DFF) and {W}avelet {F}eature {E}nhancement (WFE) module. Specifically, the DFF module uses a simple Multi-Layer Perceptron (MLP) to enhance or suppress specific frequency bands based on context signals, ensuring effective global fusion. Furthermore, the WFE module reconstructs sequences via wavelet transform, amplifying obscure non-stationary signals and short-term fluctuations that are prone to being blurred by global DFT. Finally, to ensure all frequency components are considered and to best preserve meaningful periodic user features, we blend the DFF module with the WFE module. Moreover, our proposed model achieves better performance with lower computational costs, especially in long-sequence scenarios. The main contributions of this paper are summarized as follows:

\begin{itemize}
    \item We propose a model that includes a dynamic frequency filtering module and wavelet feature enhancement module, which can efficiently fuse personalized global information with enhanced local information
    \item  Our proposed sequential recommendation model demonstrates lower computational overhead and superior recommendation performance compared to state-of-the-art baselines in long-sequence scenarios.
    \item We conducted extensive experiments on four public datasets, demonstrating the advantages of WEARec over state-of-the-art baselines.
\end{itemize}

\section{Preliminaries}

Before elaborating on the proposed WEARec, we first introduce key mathematical foundations regarding the discrete
 Fourier transform and discrete wavelet transform.
\subsection{Discrete Fourier Transform}
Given a discrete sequence $\{x_m\}_{m=0}^{n-1}$  of length $n$, it can be transformed into frequency components via:
\begin{small}
\begin{equation}
X_k = \sum_{m=0}^{n-1} x_m  e^{-2\pi imk/n} , 0 \le k \le n-1
\end{equation}
\end{small}
where $i$ denotes the imaginary unit, and $X_k$ represents the complex value of the signal at frequency index $k$. 

Simultaneously, $\{X_{k}\} ^{n-1}_{k=0}$ can be transformed back to the time-domain feature representation via the Inverse DFT (IDFT), expressed as:
\begin{small}
 \begin{equation}
 x_m= \frac{1}{n}\sum^{ n-1}_{m=0}X_{m}{e}^{2\pi imk/n}
\end{equation}
\end{small}
In our paper, we convert sequential behaviors into the frequency domain via Fast Fourier Transform (FFT) and denote it by $\mathcal{F}(·)$. Similar to IDFT, the Inverse FFT (IFFT) (denoted by $\mathcal{F}^{-1}(·)$) is also used to efficiently transfer the frequency feature back to the time domain. 

\subsection{Discrete Wavelet Transform}
Given a discrete sequence $\{x_m\}_{m=0}^{n-1}$  of length $n$, it can be decomposed into a set of high-frequency and low-frequency sub-signals through hierarchical decomposition. The $j$-th level decomposition is defined as:
\begin{small}
\begin{equation}
A_{j+1}[m] = \sum_{k=0}^{K-1} L[k]  A_j[2m-k] 
\end{equation}
\begin{equation}
D_{j+1}[m] = \sum_{k=0}^{K-1} H[k]  A_j[2m-k]
\end{equation}
\end{small}
The indexing $2m-k$ implements downsampling with stride 2, reducing output length by half. Therefore, in the equation, $K$ is $n/2^{j}$. Where $A_j[m]$ denotes the approximation coefficients after level-$j$ low-pass filtering $L$, containing the low-frequency components of the signal. When $j=0$, we set $A_{0}[m]=x[m]$. $D_{j}[m]$ denotes the detail coefficients after level-$j$ high-pass $H$ filtering, containing the high-frequency components of the signal. 
Through wavelet decomposition, Discrete Wavelet Transform (DWT) can localize transient components in time-domain signals, thereby enabling the processing and analysis of non-stationary signals.

Moreover, the decomposed high-frequency and low-frequency sub-signals can be perfectly reconstructed into the original signal via the Inverse Discrete Wavelet Transform (IDWT). It reconstructs the signal stage-by-stage via iterative upsampling and filtering operations:
\begin{small}
\begin{equation}
A_{j}[m] = \sum_{k=0}^{K-1}  \Tilde{L}[k]  A_{j+1}[2m-k] +  \sum_{k=0}^{K-1}  \Tilde{H}[k]  D_{j+1}[2m-k]
\end{equation}
\end{small}

Where $ \Tilde{L}$ and $ \Tilde{H}$ are reconstruction filters. In our paper, the forward DWT converts sequential behavior into high/low-frequency sub-signals and denote it by $\mathcal{W}(·)$. The IDWT (denoted by $\mathcal{W}^{-1}(·)$)  reconstructs decomposed sub-signals into the original signal. For more descriptions, interested readers should refer to Appendix A \cite{arxiv}.

\begin{figure*}[htbp]
\centerline{\includegraphics[scale=0.55]{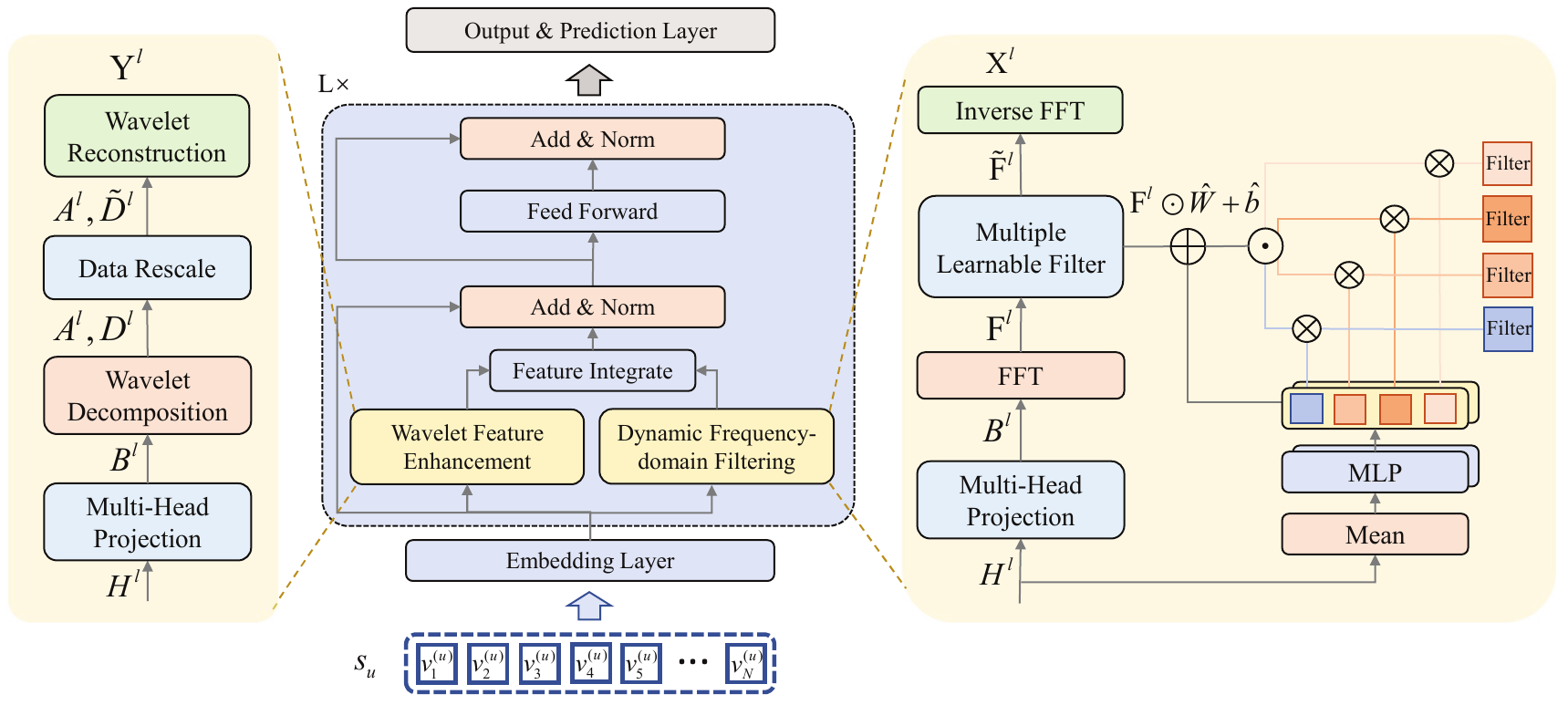}}
\caption{The model architecture of WEARec is similar to the transformer encoder. It first generates item embedding with positional embedding through the embedding layer , and then extracts user preference from the frequency domain by replacing the self-attention module with the wavelet feature enhancement module and dynamic frequency-domain filtering module. Their details are shown on both sides. Finally, a prediction layer computes a recommendation score for all candidate items.}
\label{fig.3}
\end{figure*}

\section{Proposed Method}
In this section, we first present some necessary notations to formulate the sequential recommendation problem. Additionally, we provide a comprehensive explanation of the overall framework of WEARec, as shown in Figure \ref{fig.3}.
\subsection{Problem Statement}

The goal of SR is to predict the
next item a user will click based on the user’s previous
interactions. Given a set of users $\mathcal{U}$ and items $\mathcal{V}$, where $u \in \mathcal{U}$ denotes a user and $v \in \mathcal{V}$
 denotes an item. The numbers of users and items
are denoted as $|\mathcal{U}|$ and $|\mathcal{V}|$, respectively. The set of user
behavior can be represented as $S = \{s_{1}, s_{2}, ..., s_{|\mathcal{U}|}\}$. In SR,
the user’s behavior sequence is usually in time order. This
means that each user sequence is made up of (chronologically
ordered) item interactions $s_{u} = [ v^{(u)}_1 , v^{(u)}_2 , ..., v^{(u)}_t , ..., v^{(u)}_n ]$,
where $s_u \in  S$ , $v^{(u)}_t \in \mathcal{V}$ is the item with which user
$u$ interacts at step $t$, and $n$ is the length of the sequence. Specifically, the recommendation model first divides the original sequence into multiple subsequences. After training, it generates a probability score for the candidate items in each subsequence, $i.e.$, $\hat{y} = \{ \hat{y}_1 ,\hat{y}_2 ,...,\hat{y}_{|\mathcal{V}|} \}$, where $\hat{y}_i$ denotes the prediction score of item $v_i$. Given a user’s historical interaction sequences and the maximum sequence length $N$, the sequence is first truncated by removing earliest item if $n > N$ or padded with $0$s to get a fixed length sequence $s_u = [v^{(u)}_1, v^{(u)}_2, \dots, v^{(u)}_N]$. The SR task takes
 $s_{u}$ as input to predict the top-$K$ items at the timestamp $N+1$.

\subsection{Embedding Layer}
Given a user behavior sequence $s_u$, we define the embedding representation of the sequence $\bold{E}^{u}$ using the item embedding matrix $\bold{M} \in \mathbb{R}^{|\mathcal{V}| \times d}$, where $d$ is the embedding size and $\bold{E}^{u}_{i} = \bold{M}_{s_i}$. Positional embeddings $\bold{P} \in \mathbb{R}^{N \times d}$ are used to add additional positional information while preserving the original embedding dimensionality of the items. Additionally, we perform layer normalization and dropout operations to stabilize the training process. Therefore, we generate the sequence representation $\bold{E}^{u} \in \mathbb{R}^{N \times d}$ as follows:

\begin{equation}
 \bold{E}^{u} = \mathrm{Dropout}(\mathrm{LayerNorm}(\bold{E}^{u} + \bold{P}))
\end{equation}

\subsection{Dynamic Frequency-domain Filtering}

\textbf{Multi-Head Projection.}
To enhance the representation ability of the input item embedding $\bold{E}^{u}$ in the frequency domain, we draw inspiration from the partitioning concept of the multi-head attention mechanism. Specifically, we decompose the input matrix $\bold{E}^{u} \in \mathbb{\bold{R}}^{N \times d}$ along the embedding dimension into $k$ parallel feature subspaces, each equipped with an adaptive filter tailored to its characteristics.
\begin{equation}
\bold{H}^0 = \bold{E}^{u}
\end{equation}
\begin{equation}
 \bold{H}^l = [\bold{B}_1, \bold{B}_2, \ldots, \bold{B}_k]
\end{equation}
where $\bold{H}^l \in \mathbb{R}^{N \times d}$ is the time feature of the $l$-th layer,
and $\bold{B}_i \in \mathbb{R}^{N \times d/k}$ represents the $i$-th subspace.

For each subspace, the dynamic frequency-domain filtering layer first performs a fast Fourier transform along the item dimension:
\begin{equation}
\mathcal{F}(\bold{B}_i^{l}) \to \bold{F}_i^{l}
\end{equation}
where $\bold{B}^{l}_{i} \in \mathbb{R}^{N \times d/k}$ is the $i$-th time domain subspaces feature of the $l$-th layer, and $\mathcal{F}(\cdot)$ denotes the 1D FFT. Note that $\bold{F}_{i}^{l} \in \mathbb{C}^{M \times d/k}$ is a complex tensor representing the $i$-th frequency domain subspace feature of the $l$-th layer. $M$ is calculated as:

\begin{equation}
 M = \lceil N/2 \rceil  + 1
\end{equation}

To extract the overall information of the user context sequence, we perform mean processing on the input features in the time domain along the item dimension.
\begin{small}
\begin{equation}
\bold{c}^{l} = \frac{1}{N}\sum_{i=1}^N \bold{H}_i^{l} 
\end{equation}
\end{small}
where $\bold{H}^l_{i} \in \mathbb{R}^{1 \times d}$ represents the $i$-th row of $\bold{H}^l$, and $\bold{c}^l \in \mathbb{R}^{1 \times d}$ denotes the overall representation of the user's historical interaction sequence at the $l$-th layer.

To enable dynamic adaptation of our frequency-domain filters to user-specific sequence contexts, we design two three-layer MLP networks that generate corresponding scaling factors and bias terms from captured user contextual features, thereby modulating personalized frequency-domain filters.
\begin{equation}
\bold{ \Delta s}^{l} = \mathrm{MLP_{1}}(\bold{c}^{l})
\end{equation}
\begin{equation}
\bold{ \Delta b}^{l} = \mathrm{MLP_{2}}(\bold{c}^{l})
\end{equation}
where $ \bold{ \Delta s}^{l}$ and $\bold{\Delta b}^{l}  \in \mathbb{R}^{k \times M}$ denote the scaling factor and bias term of the $l$-th layer for dynamically adjusting the filter, respectively. The scaling factor shapes the filter's overall frequency response, while the bias term adjusts weights for specific frequency bands.

Given the base filter weights $\mathbf{W}^{l} \in \mathbb{R}^{k \times M}$ and bias $\mathbf{b}^{l} \in \mathbb{R}^{k \times M}$, with $k$ representing the number of filters, the weights and bias of the personalized dynamic filter are obtained through the following operation using the personalization-generated scaling factor $\bold{ \Delta s}^{l}$ and bias term $\bold{\Delta b}^{l}$:  
\begin{equation} 
 \hat{\bold{W}}^{l} = \mathrm{\bold{W}}^{l} \odot (1 +  \bold{\Delta s}^{l})
\end{equation}
 \begin{equation}
 \hat{\bold{b}}^{l} = \mathrm{\bold{b}}^{l} + \bold{ \Delta b}^{l}
\end{equation}
where $\hat{\bold{W}}^{l}$ and $\hat{\bold{b}}^{l} \in \mathbb{R}^{k \times M}$ denote the linearly modulated weights and bias of the dynamic filter at the $l$-th layer, respectively. The modulated filter adapts to the frequency-domain characteristics of different users.

\noindent \textbf{Multiple Learnable Filter.}
By applying a linear transformation to the frequency-domain feature subspace using personalized filter weights and bias, we obtain personalized filtered frequency-domain information. 
\begin{equation}
 \Tilde{\bold{F}}_{i}^{l} = \bold{F}_{i}^{l} \odot \hat{\bold{W}}^{l} + \hat{\bold{b}}^{l}
\end{equation}

Finally, use IDFT to map the processed frequency-domain signal back to the time domain: 
\begin{equation}
 \bold{X}_{i}^{l} =  \mathcal{F}^{-1}( \Tilde{\bold{F}}_{i}^{l})
 \end{equation}

\subsection{Wavelet Feature Enhancement}
This module captures fine-grained temporal patterns through differentiable wavelet transforms. Here, the Haar wavelet transform \cite{haar} was selected due to its simple structure, high computational efficiency, and the desirable property of perfect signal reconstruction.

\noindent \textbf{Multi-Head Projection.}
To ensure alignment between the acquired fine-grained information and the spatial features obtained by dynamic frequency-domain filtering module, we extend the design philosophy of this module. 

\noindent \textbf{Wavelet Decomposition.}
 To capture fine-grained temporal patterns in behavioral sequences and enhance non-stationary signals within them, we integrate DWT into the WEARec framework. We implement Haar wavelet transform along the item dimension to decompose temporal signals into low-frequency and high-frequency components.
\begin{equation}
  \bold{A}_{i}^{l},\bold{D}_{i}^{l} = \mathcal{W}(\bold{B}_{i}^{l})
 \end{equation}
where $\bold{B}^{l}_{i} \in \mathbb{R}^{N \times d/k}$ is the $i$-th time domain subspaces feature of the $l$-th layer, and $\mathcal{W}(\cdot)$ denotes the 1D Haar wavelets transform. $\bold{A}_{i}^{l} \in \mathbb{R}^{K \times d/k}$ denotes the $i$-th subspaces' approximation coefficients representing the low-frequency components of the original signal at the $l$-th layer, while $\bold{D}_{i}^{l} \in \mathbb{R}^{K \times d/k}$ corresponds to the $i$-th subspaces' detail coefficients capturing its high-frequency components. 

\noindent \textbf{Data Rescale.}
To acquire the high-frequency information required by the model, we multiply different components of the high-frequency information by an adaptive learnable matrix, thereby adaptively enhancing or suppressing high-frequency signals in the sequence. Since low-frequency information records the original primary components of the sequence, we therefore avoid modifying it.
\begin{equation}
 \tilde{\bold{D}}_{i}^{l} = \bold{D}_{i}^{l} \odot \bold{T}^{l}
 \label{eq1}
 \end{equation}
where $\tilde{\bold{D}}_{i}^{l} \in \mathbb{R}^{K \times d/k}$ denotes the enhanced detail coefficients of the $i$-th subspaces at the $l$-th layer, enhancing the high-frequency components of the original signal, and ${\bold{T}^{l} \in \mathbb{R}^{K \times d/k}}$ denotes the adaptive high-frequency enhancer at the $l$-th layer.

\noindent \textbf{Wavelet Reconstruction.}
Finally, we reconstruct the high-frequency enhanced time-domain signal by applying the inverse Haar wavelet transform to the processed coefficients.

\begin{equation}
 \bold{Y}_{i}^{l}  = \mathcal{W}^{-1}({\bold{A}}_{i}^{l},\tilde{\bold{D}}_{i}^{l})
 \end{equation}

\subsection{Feature Integrate}
Finally, the global features extracted by the dynamic frequency filtering are mixed with the fine-grained features derived from the wavelet feature enhancement.  

\begin{equation}
\widehat{\bold{H}}^{l} = \alpha \odot \bold{X}^{l} + (1 - \alpha) \odot \bold{Y}^{l}
\end{equation}
where $\alpha$ is a hyperparameter designed to emphasize the fine-grained details enhanced by wavelet decomposition. Thus, our core design principle involves balancing wavelet-augmented local features and dynamically filtered global features.  

To prevent gradient vanishing when the model gets deeper and to achieve a more stable training process with better generalization ability, typical techniques such as skip connection, dropout, and layer normalization are implemented.
\begin{equation}
\widehat{\bold{H}}^{l}= \mathrm{LayerNorm}(\bold{H}^{l} + \mathrm{Dropout}(\widehat{\bold{H}}^{l}))
\end{equation}

\subsection{Point-wise Feed Forward Network}
To endow the models with non-linearity characteristics between different
dimensions in the time domain, we also add a feed-forward
network after each feature mixer, which consists of MLP with GELU activation. The process of the
point-wise Feed-Forward Neural network (FFN) is defined as
follows:
\begin{equation}
 \Tilde{\bold{H}}^{l} = \mathrm{FFN}(\widehat{\bold{H}}^{l})=(\mathrm{GELU}(\widehat{\bold{H}}^{l}\bold{W}_{1}+\bold{b}_{1}))\bold{W}_{2} + \bold{b}_{2}
\end{equation}
where $\bold{W}_{1},\bold{W}_{2} \in \mathbb{R}^{d \times d}$ and $\bold{b}_{1}, \bold{b}_{2} \in \mathbb{R}^{1 \times d}$
are learnable
parameters. In order to prevent overfitting, we add a dropout
layer above each hidden layer and perform layer normalization
procedures again using residual connection structure
on the output $\bold{H}^{l+1}$, as below:
\begin{equation}
\bold{H}^{l+1} = \mathrm{LayerNorm}(\widehat{\bold{H}}^{l} + \mathrm{Dropout}(\Tilde{\bold{H}}^{l}))
\end{equation}

\subsection{Prediction Layer}

In the final layer of WEARec, we can compute the recommendation probability for each candidate item to predict how likely the user would adopt the item. Specifically, the corresponding predicted probability
$\hat{\bold{y}}$ can be generated by:
\begin{equation}
\hat{\bold{y}} = \mathrm{softmax}(\bold{h}^{L}(\bold{M})^{\top})
\end{equation}
where $\hat{\bold{y}} \in \mathbb{R}^{|\mathcal{V}|}$. and $\bold{h}^{L} \in \mathbb{R}^{1 \times d}$ is the output of the $L$-layer blocks at the final step. To optimize the model parameters, we therefore use the cross-entropy loss \cite{duo,fea,slime,bsa}.
The objective function of SR can be formulated as:
\begin{small}
\begin{equation}
\mathcal{L}_{Rec} = -\sum_{i=1}^{|\mathcal{V}|}y_{i}\mathrm{log}(\hat{y}_{i})
\end{equation}
\end{small}
where $y_{i}$ is the $i$-th ground truth item, and $\hat{y}_{i}$ denotes the preference score of $v_i$.
\section{Experiments}
In this section, we first briefly introduce the datasets used in our experiments, nine baselines, the evaluation metrics, and
the implementation details in our experimental settings. Then,
we compare our proposed model WEARec with state-of-the-art baseline methods. Specifically, to study the validity of WEARec, we conduct
experiments to try to answer the following questions:

\textbf{RQ1} Does WEARec perform better than the state-of-
the-art baselines?

\textbf{RQ2} How does WEARec perform and what is its computational overhead in long-sequence scenarios?

\textbf{RQ3} How does each designed module in WEARec contribute to the performance?

\textbf{RQ4} How do the hyper-parameters affect the effectiveness of WEARec?
 \subsection{Experimental Setup}
 
\noindent \textbf{Datasets.} We conduct experiments on four public datasets collected from real-world platforms in order to thoroughly evaluate WEARec.
 i,ii) Beauty and Sports from Amazon \cite{amazon}, iii) ML-1M \cite{movielens}, iv) LastFM. Following \cite{s3,fmlp} , we also adopt the 5-core settings by filtering out users with less than 5 interactions. The detailed dataset statistics are presented in Appendix B.1 \cite{arxiv}.

\noindent \textbf{Evaluation Metrics.} In our evaluation, we adopt the leave-one-out strategy for partitioning each user’s
 item sequence \cite{s3}. We rank
the prediction scores throughout the entire item set without
using negative sampling, as recommended by \cite{sampled}.
Performance is evaluated on a variety of evaluation metrics, including Hit
Ratio at $K$ (HR@$K$) and Normalized Discounted Cumulative
Gain at $K$ (NDCG@$K$, NG@$K$) on all datasets. The $K$ is set to 10 and 20.

\begin{table*}[t]
\begin{center}
\setlength{\tabcolsep}{0.7mm}
\small
{
\begin{tabular}{l|l|ccccccccccc}
\toprule
{Datasets} & {  Metric} & 
{Caser} & {GRU4Rec} & {SASRec} & {DuoRec} & {FMLPRec} & {FamouSRec}  & 
{FEARec}  & {SLIME4Rec} & {BSARec} & \textbf{WEARec} & {Improv.}\\
\midrule

\multirow{4}{*}{Beauty} 
& HR@10 & 0.0225 & 0.0304 & 0.0531  & 0.0965& 0.0559 & 0.0838  & 0.0982 & 0.1006&  \underline{0.1008}&\textbf{0.1041} & 3.27\% \\
& HR@20 & 0.0403 & 0.0527 & 0.0823  & 0.1313 & 0.0869 & 0.1146  & 0.1352  & \underline{0.1381}& {0.1373} & \textbf{0.1391} & 1.31\% \\
& NG@10 & 0.0108 & 0.0147 & 0.0283& 0.0584  & 0.0291 & 0.0497& 0.0601 & 0.0601& \underline{0.0611}& \textbf{0.0614} & 0.49\%  \\
& NG@20 & 0.0153 & 0.0203 & 0.0356 & 0.0671  & 0.0369 & 0.0575 & 0.0694  & 0.0696 & \textbf{0.0703}& \textbf{0.0703} & 0.00\% \\
\midrule

\multirow{4}{*}{Sports} 
& HR@10 & 0.0163 & 0.0187 & 0.0298 & 0.0569  & 0.0336 & 0.0424 & 0.0589  & 0.0611 & \underline{0.0612}& \textbf{0.0631}  & 3.10\% \\
& HR@20 & 0.0260 & 0.0303 & 0.0459& 0.0791  & 0.0525 & 0.0632  & 0.0836 & \underline{0.0869} & {0.0858} & \textbf{0.0895} & 2.99\%  \\
& NG@10 & 0.0080 & 0.0101 & 0.0159  & 0.0331 & 0.0183 & 0.0244 & 0.0343  & 0.0357 & \underline{0.0360} & \textbf{0.0367} & 1.94\%  \\
& NG@20 & 0.0104 & 0.0131 & 0.0200 & 0.0387 & 0.0231 & 0.0297  & 0.0405  & 0.0421 & \underline{0.0422}& \textbf{0.0433}  & 2.60\% \\
\midrule

\multirow{4}{*}{LastFM} 
& HR@10 & 0.0431 & 0.0404 & 0.0633  & 0.0624 & 0.0560 & 0.0569 & 0.0587 & 0.0633 & \underline{0.0807}& \textbf{0.0899}  & 11.40\%  \\
& HR@20 & 0.0642 & 0.0541 & 0.0927  & 0.0963 & 0.0826 & 0.0954  & 0.0826 & 0.0927 & \underline{0.1174}&\textbf{0.1202}  & 2.38\% \\
& NG@10 & 0.0268 & 0.0245 & 0.0355 & 0.0361& 0.0306 & 0.0318  & 0.0354& 0.0359 & \underline{0.0435}& \textbf{0.0465}  & 6.89\% \\
& NG@20 & 0.0321 & 0.0280 & 0.0429 & 0.0446 & 0.0372 & 0.0415 & 0.0414 & 0.0433& \underline{0.0526} & \textbf{0.0543}  & 3.23\%  \\
\midrule

\multirow{4}{*}{ML-1M} 
& HR@10 & 0.1556 & 0.1657 & 0.2137& 0.2704 & 0.2065 & 0.2639  & 0.2705 & \underline{0.2891} & {0.2757}& \textbf{0.2952}  & 2.10\% \\
& HR@20 & 0.2488 & 0.2664 & 0.3245 & 0.3738 & 0.3137 & 0.3717  & 0.3714 & \underline{0.3950}& {0.3884}&  \textbf{0.4031} & 2.05\% \\
& NG@10 & 0.0795 & 0.0828 & 0.1116 & 0.1530 & 0.1087 & 0.1455 & 0.1516 & \underline{0.1673} & {0.1568}& \textbf{0.1696} & 1.37\% \\
& NG@20 & 0.1028 & 0.1081 & 0.1395 & 0.1790  & 0.1356 & 0.1727  & 0.1771  & \underline{0.1939}& {0.1851}& \textbf{0.1968} & 1.49\% \\
\bottomrule
\end{tabular}}
\end{center}
\caption{Recommendation algorithms performance comparison on 4 datasets. The best results are in boldface and the second-best results are underlined. ‘Improv.' indicates the relative improvement against the best baseline performance.}
\label{tab1}
\end{table*}

\noindent \textbf{Baseline Models.}
To demonstrate the effectiveness of the proposed model, we
compare WEARec with the most widely used and state-of-the-art methods with two categories:

\textbf{Time-domain SR models}: GRU4Rec \cite{gru}, Caser \cite{caser}, SASRec \cite{sasrec}, and DuoRec \cite{duo}.

\textbf{Frequency-domain SR models}: FMLPRec \cite{fmlp}, FamouSRec \cite{famous}, FEARec \cite{fea}, SLIME4Rec \cite{slime}, BSARec \cite{bsa}.

\noindent \textbf{Implementation Details.}
We implement our WEARec model in PyTorch. For the baseline models, we refer to their best hyper-parameters setups reported in
the original papers and directly report their reimplementations
results if available, since the datasets and evaluation metrics
used in these works are strictly consistent with ours. Both the dimension of the feed-forward network and item embedding size are set to 64. The number of WEARec blocks $L$ is set to 2, and the maximum sequence length $N$ is set to 50. Batch size is set to 256. The model is optimized by Adam optimizer with a learning rate  from \{0.0005,0.001\}. The wavelet decomposition level is set to 1. The $\alpha$ is in \{0.1,0.2,0.3,0.4,0.5,0.6,0.7,0.8,0.9\}, and the $k$ chosen from \{1,2,4,8\}. We report the result of each model under its optimal hyper-parameter settings. The best hyperparameters are in Appendix B.3 \cite{arxiv} for reproducibility.

\subsection{Recommendation Performance Comparison (RQ1)}
The overall experimental results on four datasets are presented in Table \ref{tab1}. Based on these results, we can draw the following observations and conclusions. Firstly, traditional time-domain-based sequential recommendation methods, such as Caser, GRU4Rec, and SASRec, exhibit suboptimal performance. This is because they fail to adequately identify intertwined user periodic patterns, which are crucial for capturing users' true interests. DuoRec validates the effectiveness of combining supervised and unsupervised contrastive learning through model and semantic augmentation. Secondly, among these models, methods leveraging the frequency domain ($e.g.$, FMLPRec, FamouSRec, FEARec, SLIME4Rec, BSARec) generally demonstrate superior performance. FMLPRec, by utilizing an MLP structure to attenuate noise in the frequency domain, achieved nearly comparable or even better performance than SASRec on most datasets. FamouSRec, FEARec, and SLIME4Rec further advanced this direction by combining frequency-domain analysis with contrastive learning, achieving better performance. BSARec mitigated the insufficient inductive bias of the self-attention mechanism, enhanced the performance of the attention mechanism, and alleviated over-smoothing through a frequency recalibrator. Finally, based on these results, WEARec achieved the best performance across all four datasets by combining a dynamic frequency-domain filtering module with wavelet feature enhancement module.
\subsection{Model in Long Sequence Scenarios (RQ2)}

Given the sparsity of most datasets in recommendation systems, the maximum sequence length $N$ is often limited to 50 during the evaluation of sequential recommendation models. However, this setting is not appropriate for relatively dense datasets with frequent user interactions.
\begin{figure}[tbp]

\centering
\centerline{\includegraphics[width=0.48\textwidth]{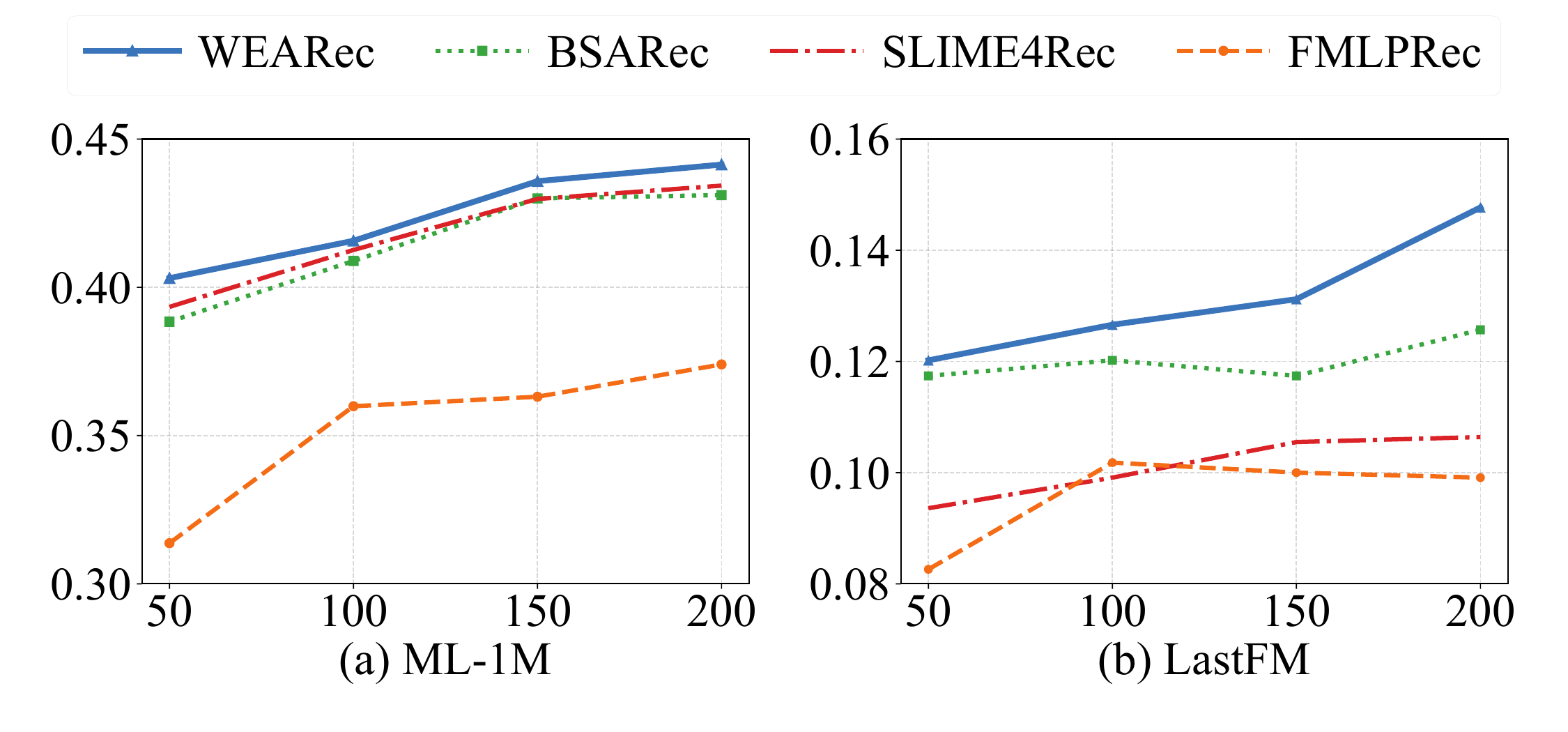}}
\caption{The HR@20 performance comparison of WEARec with FMLPRec, SLIME4Rec and BSARec at different sequence lengths $N$ on  ML-1M and LastFM.}
\label{Fig.3}
\end{figure}

\begin{table}[t]
\begin{center}
\renewcommand{\arraystretch}{1.3}
\setlength{\tabcolsep}{1mm}
{
\begin{tabular}{c c c c c c c c  c }
\toprule
\multicolumn{1}{c}{\multirow{2}{*}{Methods}} & \multicolumn{2}{c}{ML-1M} & \multicolumn{2}{c}{LastFM}\\ \cline{2-5}
           &\# params & s/epoch &\# params & s/epoch \\
\hline
WEARec &{426,082} &{66.46} &{440,802} &{5.23} \\
\hline
FMLPRec &{324,160} &{36.93}  &{338,880} &{4.91}\\
BSARec &{331,968} &{109.26}&{346,688} &{10.84}\\
SLIME4Rec &{375,872} &{120.43}  &{390,592} &{13.77}\\
\bottomrule
\end{tabular}
}
\end{center}
\caption{The number of parameters and training time (runtime per epoch) for $N=200$ on  ML-1M and LastFM. More results are in Appendix B \cite{arxiv}.}
\label{tab4}
\end{table}

\noindent \textbf{Model performance.}
To investigate the impact of long sequence scenarios on recommendation results, we varied the maximum sequence length $N$ for FMLPRec, BSARec, SLIME4Rec and WEARec. We selected the LastFM and ML-1M datasets, which have longer average sequence lengths, for our experiments. Figure \ref{Fig.3} presents the experimental results in terms of HR@20. We have obtained similar experimental results in terms of other metrics. We observed that almost all models achieved their best performance at $N=200$, indicating that longer sequence information can more comprehensively represent user behavior patterns. Furthermore, while baseline models showed performance improvements in long-sequence scenarios, they were prone to overfitting, leading to performance convergence. Finally, WEARec consistently outperformed the baselines across all different maximum sequence length settings, and its improvement over baseline models was even more significant in long-sequence scenarios. For more descriptions, interested readers should refer to Appendix B.4 \cite{arxiv}.

\noindent \textbf{Model Complexity and Runtime Analyses.} To evaluate the overhead of WEARec, we assessed the number of parameters and runtime per epoch during training  at $N=200$. The results are presented in Table \ref{tab4}. We can observe that WEARec exhibits a shorter training time compared to baseline models with competitive performance. Overall, WEARec's total parameters are increased due to the use of a simple MLP. However, by not employing contrastive learning and self-attention mechanisms, WEARec actually trains faster than SLIME4Rec and BSARec. For more descriptions, interested readers should refer to Appendix B.5 \cite{arxiv}.

\subsection{In-depth Model Analysis (RQ3-RQ4)}

\begin{figure}[tbp]
\centering
\subfigure[HR@20]{
\label{Fig4.sub.1}
\includegraphics[width=0.22\textwidth]{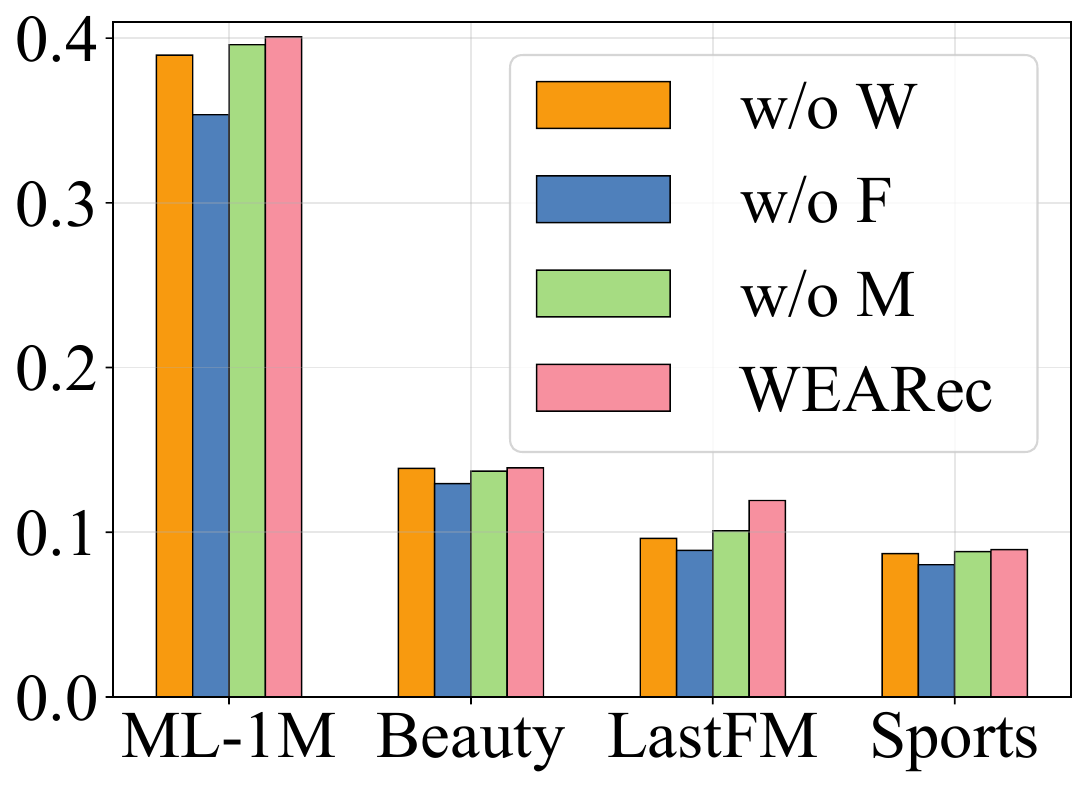}}
\subfigure[NG@20]{
\label{Fig4.sub.2}

\includegraphics[width=0.22\textwidth]{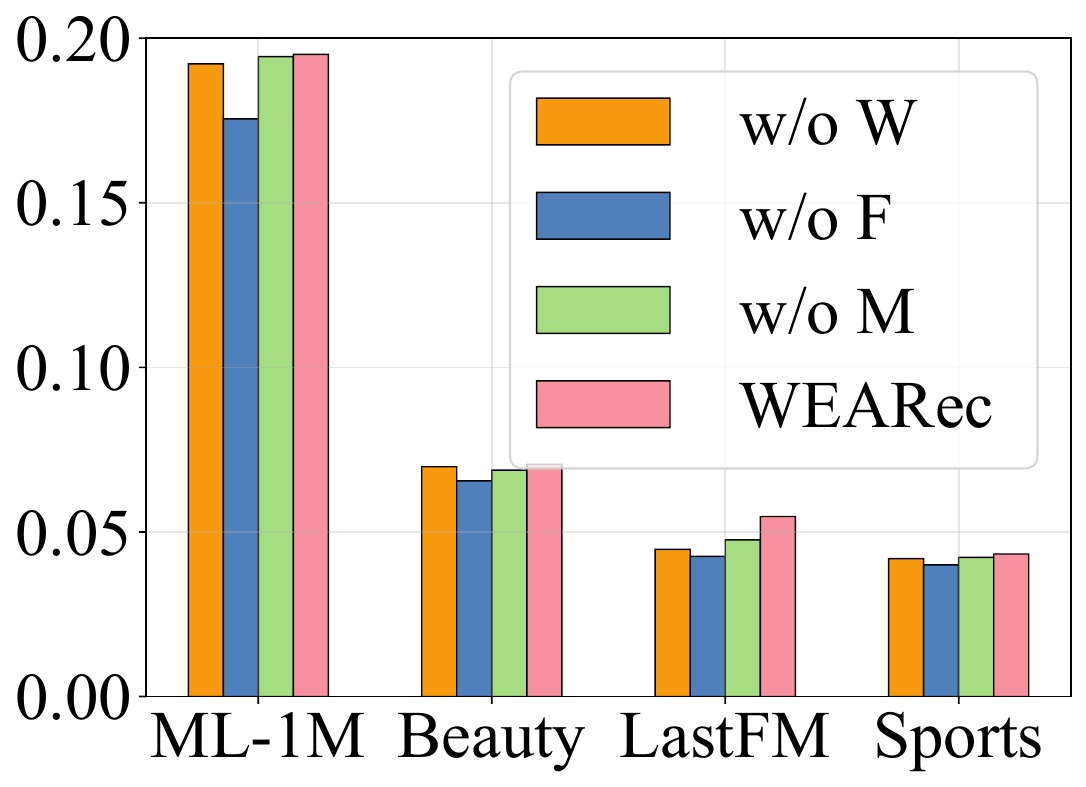}}
\caption{The HR@20 and NG@20 performance achieved by WEARec variants on four datasets.}
\label{Fig4}
\end{figure}

\noindent \textbf{Ablation study (RQ3).} Figure \ref{Fig4} summarizes the HR@20 and NG@20 performance of WEARec and its variants across four datasets. In this figure, WEARec represents the full WEARec model, while $w/o$ W, $w/o$ F and $w/o$ M represent variants where the WFE module, DFF module, and multi-head projection are removed, respectively, with all other components remaining unchanged. The results show that WEARec outperforms its variants on all datasets, indicating that all components are effective.

\noindent \textbf{Hyper-parameter analysis (RQ4).}

\begin{figure}[tbp]

\centering
\centerline{\includegraphics[width=0.48\textwidth]{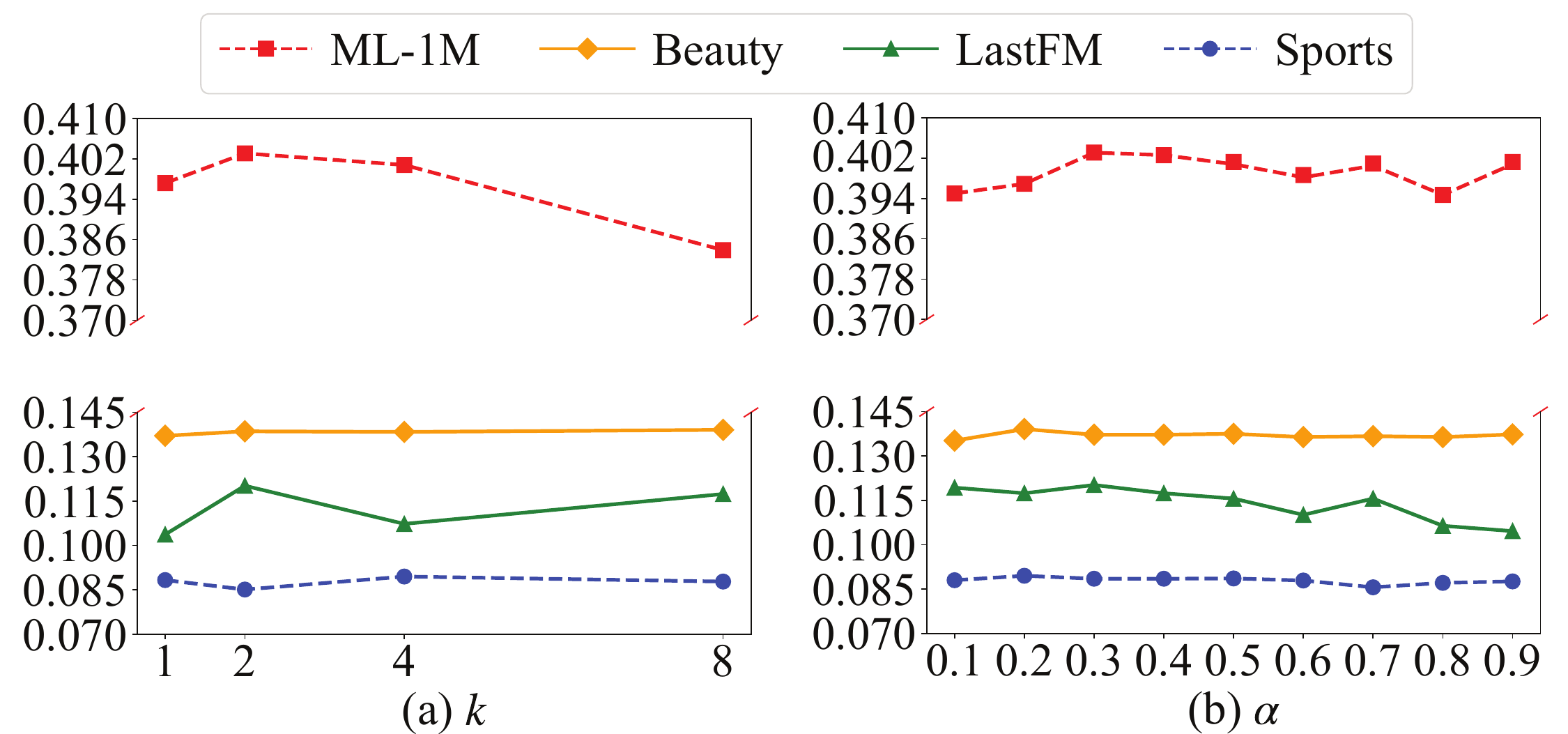}}
\caption{Performance of WEARec on HR@20 with varying
 hyperparameters..}
\label{Fig.5}
\end{figure}

\noindent \textbf{Sensitivity to $k$.}
Figure \ref{Fig.5} shows the HR@20 by varying $k$. The results indicate that an optimal $k$ value, neither too large nor too small, is critical for learning user interest preferences and consequently improving model performance.

\noindent \textbf{Sensitivity to $\alpha$.}
Figure \ref{Fig.5} shows the HR@20 by varying $\alpha$. These results suggest that optimal performance is more probable when $\alpha$ is approximately 0.3.

\noindent \textbf{Visualization of the filters.}
\begin{figure}[tbp]
\centering
\centerline{\includegraphics[width=0.48\textwidth]{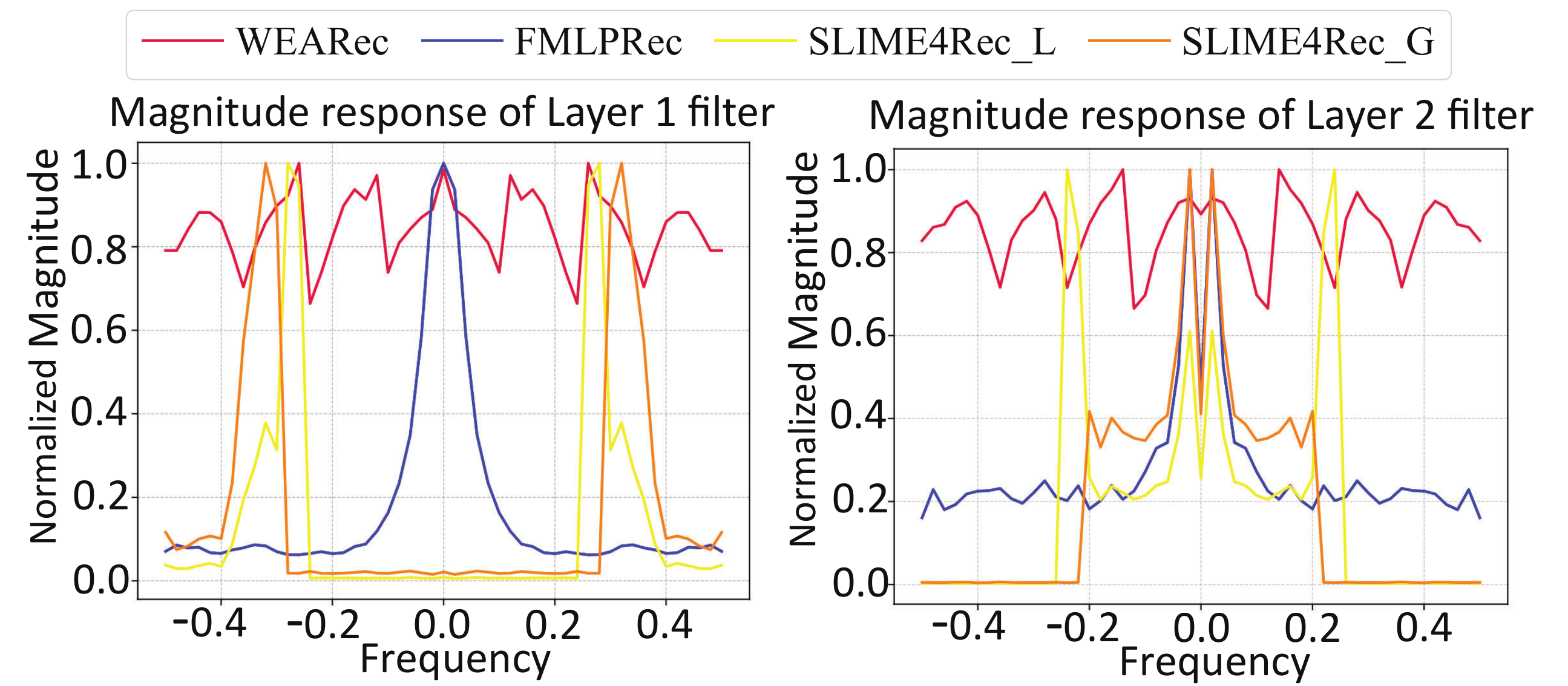}}
\caption{Visualization of spectral responses for different types of filter models across layers in Beauty. More
in-depth model analysis in Appendix C \cite{arxiv}.}
\label{Fig.6}
\end{figure}
Figure \ref{Fig.6} presents the frequency and amplitude features learned by different types of filtering models. Due to static filter design, both FMLPRec and SLIME4Rec tend to learn low-frequency components within their respective frequency bands. Conversely, WEARec, benefiting from its dynamic frequency-domain filtering design, is capable of encompassing all frequency components.

\section{Related Works}
\subsection{Time-domain SR Models}
Early SR research often relied on Markov chain assumptions \cite{fmc}. With the widespread adoption of deep learning methods  \cite{ncf}, numerous studies have employed neural network architectures as encoders. Caser \cite{caser} utilizes convolutional operations to capture higher-order patterns. SASRec \cite{sasrec} leverages self-attention mechanisms to capture item-item relationships. Furthermore, recent studies have enhanced sequential embedding representations through contrastive learning ($e.g.$, CL4SRec \cite{cl4rec} and DuoRec \cite{duo}). However, these time-domain models still struggle to effectively capture users' underlying periodic behavioral patterns.
\subsection{Frequency-domain SR Models}
Recently, researchers have begun applying frequency-domain analysis to sequential recommendation. FMLPRec \cite{fmlp} pioneered frequency-based MLP filtering to capture periodic patterns. SLIME4Rec \cite{slime} and FEARec \cite{fea} further advanced this direction by proposing a layered frequency ramp structure integrated with contrastive learning. BSARec \cite{bsa} sought to uncover fine-grained sequential patterns and inject them as inductive biases into the model. FamouSRec  \cite{famous}  developed a MoE architecture for selecting specialized expert models tailored to users' specific frequency-based behavioral patterns. However, these models either lack user-specific adaptivity or incur substantial computational costs.
\section{Conclusion}
In this paper, we introduce WEARec, a more efficient model for handling long sequences in sequential recommendation tasks, designed to effectively capture diverse user behavioral patterns. Our method includes dynamic frequency-domain filtering and wavelet feature enhancement. The former dynamically adjusts filters based on user sequences to obtain personalized frequency-domain global distributions. The latter reconstructs sequences through wavelet transforms to enhance non-stationary signals. Extensive experiments on four public datasets validate the effectiveness of WEARec.
\section{Acknowledgements}
 This research was partially supported by the NSFC
 (62376180, 62176175, 62572335), the National Key Research and
 Development Program of China (2023YFF0725002),
 Suzhou Science and Technology Development Program
 (SYG202328), and the Priority Academic Program Develop
ment of Jiangsu Higher Education Institutions.

\bibliography{aaai2026}
\appendix
\section{Appendix}
\subsection{A More Preliminaries}
\subsection{A.1 Discrete Fourier Transform}
The DFT is a core mathematical tool in signal processing,
used to convert discrete-time signals from the time domain
to the frequency domain, revealing the frequency components and energy distribution of signals. 
The key to understanding DFT lies in its two core theorems: Parseval's theorem and the convolution theorem.
\subsubsection{Parseval's theorem.} It establishes the conservation of a signal's total energy before and after the transform. For an input sequence $\{x_m\}_{m=0}^{n-1}$ and its Fourier transform $\{X_{k}\} ^{n-1}_{k=0}$ ,the total signal energy remains unchanged, apart from a constant scaling factor. That energy conservation is crucial for our method. It guarantees that adaptive filtering won't accidentally distort the intrinsic information of the input signal. This theorem can be expressed as:
 \begin{equation}
\sum_{m=0}^{n-1} |x[m]|^2 = \frac{1}{n} \sum_{k=0}^{n-1} |X[k]|^2
\end{equation}

\subsubsection{The convolution theorem.} It reveals that time-domain convolution is equivalent to element-wise multiplication in the frequency domain. This property directly enables the design of filters that capture specific frequency components.

Given the input sequence signal $\{x_m\}_{m=0}^{n-1}$ and convolution parameters $\{h_m\}_{m=0}^{n-1}$, the discrete convolution operation yields:

\begin{equation}
h[m] \ast x[m] = \sum_{m=0}^{n-1} h[m] x_{n}[m-b]
\end{equation}

Where $ \ast $ denotes the circular convolution operator, and $x_{n}[\cdot]$ is the period extension of $\{x_m\}_{m=0}^{n-1}$.
The convolution theorem proves that, given the transformed frequency feature $X[k]$, the above equation can be transformed into:
\begin{equation}
h[m] \ast x[m] = \mathcal{F}^{-1}(\mathcal{F}(h[m]) \odot X[k])
\end{equation}

Where $F(h[m])$ is the learnable filter, and $\odot$ denotes the hadamard product (element-wise multiplication). This formula can be simplified to:
\begin{equation}
\mathcal{F}(h \ast x) = \bold{W} \odot \bold{X}
\end{equation}

Where $\bold{W} \in \mathbb{C}$ corresponds to the learnable complex-valued filtering matrix, and $\bold{X} \in \mathbb{C}$ represents the frequency-domain representation of the input signal.
\subsection{A.2 Discrete Wavelet Transform}
Discrete Wavelet Transform (DWT) is a multiresolution analysis-based signal processing tool that achieves time-frequency localized decomposition of signals via scaling and shifting operations on wavelet basis functions. 

The properties of a wavelet transform depend on the choice of wavelet. For each wavelet type, there exists a father wavelet function and a mother wavelet function, used in wavelet transforms to extract the overall trend characteristics and detailed features of signals, respectively. In DWT, these functions are replaced by a set of discrete filters $L$ and $H$. $L$ represents the low-pass filter, extracting low-frequency information ($e.g.$, the signal’s overall trend), corresponding to the father scaling function. $H$ represents the high-pass filter, extracting high-frequency information ($e.g.$, the signal’s detailed features), corresponding to the mother wavelet function. They satisfy the relation: $H[n]=(-1)^{n}L[1-n]$.

\begin{table}[t]
\begin{center}
\renewcommand{\arraystretch}{1.3}
{
\begin{tabular}{c c c c c c}
\toprule
Specs. &{LastFM} &{ML-1M}  &{Beauty} &{Sports}  \\
\hline
\# User &{1,090}  &{6,041} &{22,363}   &{25,598}\\
\# Items &{3,646} &{3,417} &{12,101}  &{18357} \\
\# Interactions  &{52,551} &{999,611} &{198,502} &{296,337}\\
\# Avg.Length&{48.2} &{165.5} &{8.9} &{8.3}\\
Sparsity &{98.68\%} &{95.16\%}  &{99.93\%} &{99.95\%}\\
\bottomrule
\end{tabular}
}
\end{center}
\caption{Statistics of the datasets after preprocessing.}
\label{tabdataset}
\end{table}
\begin{table}[t]
\begin{center}
\renewcommand{\arraystretch}{1.2}
{
\begin{tabular}{c c c c c c}
\toprule
Specs. &{LastFM} &{ML-1M}  &{Beauty} &{Sports}  \\
\hline
$\alpha$ &{0.3}  &{0.3} &{0.2}   &{0.3} \\
$k$ &{2} &{2} &{8}  &{4} \\
lr  &{0.001} &{0.0005} &{0.0005} &{0.001}\\
\bottomrule
\end{tabular}
}
\end{center}
\caption{Best hyperparameters of WEARec on all datasets.}
\label{tabparam}
\end{table}
\subsection{B Additional Details for Experiments}

\begin{table*}[t] 
\begin{center}
\setlength{\tabcolsep}{1.5mm}
{
    \begin{tabular}{ll cccc cccc}
    \toprule
    \multirow{2}{*}{Method} & & \multicolumn{4}{c}{ML-1M} & \multicolumn{4}{c}{LastFM} \\
    \cmidrule(lr){3-6} \cmidrule(lr){7-10}
    & & HR@10 & NG@10 & HR@20 & NG@20 & HR@10 & NG@10 & HR@20 & NG@20 \\
    \midrule
    \multirow{3}{*}{$N=50$} & BSARec & 0.2757 & 0.1568 & 0.3884 & 0.1851 & 0.0807 & 0.0435 & 0.1174 & 0.0526 \\
    & SLIME4Rec & 0.2894 & 0.1675 & 0.3934 & 0.1937 & 0.0633 & 0.0376 & 0.0936 & 0.0453 \\ 
    &\textbf{Ours}& \textbf{0.2952} & \textbf{0.1696} & \textbf{0.4031} & \textbf{0.1968} & \textbf{0.0899} & \textbf{0.0465} & \textbf{0.1202} & \textbf{0.0547} \\

    \midrule
    
    \multirow{3}{*}{$N=100$} & BSARec & 0.3073 & 0.1815 & 0.4089 & 0.2024 & 0.0798 & 0.0455 & 0.1202 & 0.0545 \\ & SLIME4Rec & 0.3147 & 0.1815 & 0.4126 & 0.2062 & 0.0679 & 0.0382 & 0.0991 & 0.0463 \\
    & \textbf{Ours} & \textbf{0.3180} & \textbf{0.1819} & \textbf{0.4175} & \textbf{0.2069} & \textbf{0.0890} & \textbf{0.0494} & \textbf{0.1266} & \textbf{0.0589} \\

    \midrule
    
    \multirow{3}{*}{$N=150$} & BSARec & 0.3171 & 0.1826 & 0.4300 & 0.2111 & 0.0826 & 0.0476 & 0.1174 & 0.0564 \\ & SLIME4Rec & 0.3166 & 0.1820 & 0.4298 & 0.2127 & 0.0688 & 0.0387 & 0.1055 & 0.0479 \\
   & \textbf{Ours}& \textbf{0.3215} & \textbf{0.1848} & \textbf{0.4338} & \textbf{0.2131} & \textbf{0.0927} & \textbf{0.0522} & \textbf{0.1312} & \textbf{0.0617} \\

        \midrule
    \multirow{3}{*}{$N=200$} & BSARec & 0.3161 & 0.1837 & 0.4311 & 0.2127 & 0.0862 & 0.0476  & 0.1257 & 0.0594 \\ & SLIME4Rec & 0.3166 & 0.1850 & 0.4343 & 0.2173 & 0.0679 & 0.0391 & 0.1064 & 0.0488 \\
    & \textbf{Ours} & \textbf{0.3334} & \textbf{0.1904} & \textbf{0.4421} & \textbf{0.2179} & \textbf{0.0972} & \textbf{0.0556} & \textbf{0.1477} & \textbf{0.0682} \\

    \bottomrule
    \end{tabular}}
\end{center}
  \caption{Performance comparison of WEARec with SLIME4Rec and BSARec at different sequence lengths $N$. HR is the abbreviation for Hit Ratio, and NG is the abbreviation for NDCG. The best performing part in each row is shown in bold.}
  \label{tab:performance}
\end{table*}

\begin{table*}[t]
\begin{center}
\renewcommand{\arraystretch}{1.4}
{
\begin{tabular}{c c c c c c c c  c }
\toprule
\multicolumn{1}{c}{\multirow{2}{*}{Methods}} & \multicolumn{2}{c}{ML-1M}& \multicolumn{2}{c}{Beauty}& \multicolumn{2}{c}{Sports}& \multicolumn{2}{c}{LastFM}\\ \cline{2-9}
           &\# params & s/epoch &\# params & s/epoch&\# params & s/epoch &\# params & s/epoch \\
\hline
WEARec &{426,082} &{66.46} &{981,922} &{15.12}&{1,382,306} &{26.12} &{440,802} &{5.23} \\
\hline
FMLPRec &{324,160} &{36.93} &{880,000} &{10.11}&{1,280,384} &{22.78} &{338,880} &{4.91}\\
BSARec &{331,968} &{109.26} &{887,808} &{25.87}&{1,288,192} &{50.59} &{346,688} &{10.84}\\
SLIME4Rec &{375,872} &{120.43} &{931,712} &{31.44}&{1,332,096} &{68.74} &{390,592} &{13.77}\\
\bottomrule
\end{tabular}
}

\end{center}
\caption{The number of parameters and training time (runtime per epoch) for $N=200$ on all datasets.}
\label{tabtime}
\end{table*}

\subsection{B.1 Details of Datasets}
We provide 4 benchmark datasets used for our experiments. These datasets, which differ in scenarios, sizes, and sparsity, are frequently used in tests of sequential recommendation methods. The main statistics of four datasets after preprocessing are reported in Table \ref{tabdataset}. We elaborate on the descriptions of the individual dataset below.
\begin{itemize}
    \item \textbf{LastFM} contains user interaction with music, such as
artist listening records. It is used to recommend musicians to users in sequential recommendation with long sequence lengths.
    \item  \textbf{MovieLens-1M} \cite{movielens} is based on reviews of movies that
were collected from the non-commercial movie recommendation website MovieLens. The interaction number
in ML-1M is about 1 million.
    \item  \textbf{Amazon Beauty, and Sports} \cite{amazon} These datasets
    contain user-item interactions from the Amazon review dataset. They consist of product reviews and ratings, providing a rich data source for evaluating sequential recommendation models
in the e-commerce domain.
\end{itemize}
\subsection{B.2 Details of Baselines}
To demonstrate the effectiveness of the
 proposed model, we compare WEARec with the most
 widely used and state-of-the-art methods with two cate
gories: 
\begin{itemize}
\item \textbf{Time-domain SR models}: \textbf{GRU4Rec} \cite{gru} is the first model to apply Gated Recurrent Unit (GRU) to model sequences of user behavior for sequential recommendation, \textbf{Caser} \cite{caser} is a CNN-based method capturing local dynamic patterns of user activity by using horizontal and vertical
convolutional filters over time, \textbf{SASRec} \cite{sasrec} captures relations between items in a sequence using the self-attention mechanism, and \textbf{DuoRec} \cite{duo} uses unsupervised model-level augmentation and supervised semantic positive samples for contrastive learning. 

\item \textbf{Frequency-domain SR models}: \textbf{FMLPRec} \cite{fmlp} is a all-MLP model using a learnable filter enchanced block to remove noise in the embedding matrix, \textbf{FamouSRec} \cite{famous} uses a Mixture-of-Experts (MoE) approach, allowing the
model to focus on different frequency ranges via heterogeneous encoder modules, \textbf{FEARec} \cite{fea} utilizes frequency domain information in atten-
tion computation and integrates information from both time
and frequency domains, \textbf{SLIME4Rec} \cite{slime} utilizes a frequency ramp structure to consider different frequency bands for each layer with dynamic and
static selection modules, \textbf{BSARec} \cite{bsa} adjusts the influence on the high-frequency region to be learnable and utilizes it as an inductive bias of self-attention. It is the most recent and strong baseline for sequential
recommendation. 
\end{itemize}

\subsection{B.3 Experimental Settings \& Hyperparameters}

The following software and hardware environments were
used for all experiments: PYTHON 3.9.7, PYTORCH 2.3.0, NUMPY 1.24.3, SCIPY 1.11.1,
CUDA 12.2, and NVIDIA Driver 535.104.05, and Intel Xeon Gold 6248R CPU, and 40GB NVIDIA Tesla A100 GPU.
For reproducibility, we introduce the best hyperparameter configurations for each dataset in Table \ref{tabparam}. We conducted
experiments under the following hyperparameters: the $\alpha$ is
in \{0.1, 0.2, 0.3, 0.4, 0.5, 0.6, 0.7, 0.8, 0.9\}, and the number of filters $k$ is chosen from \{1, 2, 4, 8\}. For training, the Adam optimizer is optimized with learning rate in \{0.0005, 0.001\}.Moreover, to address the sparsity of Amazon datasets and LastFM dataset, a dropout rate of 0.5 is used, compared to 0.1 for MovieLens-1M.
\subsection{B.4 Model Performance}
To investigate the impact of long sequence scenarios on recommendation results, we varied the maximum sequence length $N$ for FMLPRec, BSARec, SLIME4Rec and WEARec. Tabel \ref{tab:performance} presents the experimental results. We observed that almost all models achieved their best performance at $N=200$, indicating that longer sequence information can more comprehensively represent user behavior patterns. Furthermore, while baseline models showed performance improvements in long-sequence scenarios, they were prone to overfitting, leading to performance convergence. Finally, WEARec consistently outperformed the baselines across all different maximum sequence length settings, and its improvement over baseline models was even more significant in long-sequence scenarios.
\subsection{B.5 Model Complexity and Runtime Analyses}

Different from transformer-based models that rely on the
self-attention mechanism, WEARec is an attention-free
architecture with dynamic frequency-domain filtering module and wavelet feature enhancement module. The computation
complexity of traditional self-attention is $\mathcal{O}(n^{2}d+nd^{2})$, where
$n$ is the sequence length of user and $d$ is hidden size. In contrast, the time complexity of the feature mixing layer can be reduced to $\mathcal{O}(nd \mathrm{log}n + nd)$ with FFT and Haar wavelets transform. For the adaptive filtering, which includes two MLPs for context-based parameter generation, the time complexity is $\mathcal{O}(2nd^{2})$, and the time cost of feed-forward networks is $O(nd^{2})$. Therefore, the total time complexity of WEARec is $\mathcal{O}((nd \mathrm{log} n + nd + 3{nd^{2}}))$ which is proportional to the log-linear complexity of the input
sequence length $n$.
Because $\mathcal{O}(nd \mathrm{log}n)$ typically dominates the cost of element-wise operations
for large $n$, the total complexity remains $\mathcal{O}(nd \mathrm{log}n)$. This demonstrates that our model exhibits superior time complexity compared to models using self-attention mechanisms with $\mathcal{O}(n^{2})$ overhead ($e.g.$, BSARec). Furthermore, as our model does not employ contrastive learning, it also achieves lower computational overhead than sequential recommendation models leveraging contrastive learning ($e.g.$, SLIME4Rec).   

To evaluate the complexity and efficiency of WEARec, we evaluate the number of parameters and runtime per epoch. Results for all datasets are shown in Table \ref{tabtime}. Overall, WEARec increases the total parameters. We show that WEARec has faster runtime times per epoch than FEARec and DuoRec across all datasets.

\subsection{C More Visualization and Case Study }
\subsection{C.1 Visualization of Learned $\bold{T}$}
\begin{figure}[t]
	\centering  
	\subfigbottomskip=2pt 
	\subfigcapskip=-5pt 
	\subfigure[WEARec,WFE,1 layer]{
		\includegraphics[width=0.225\textwidth]{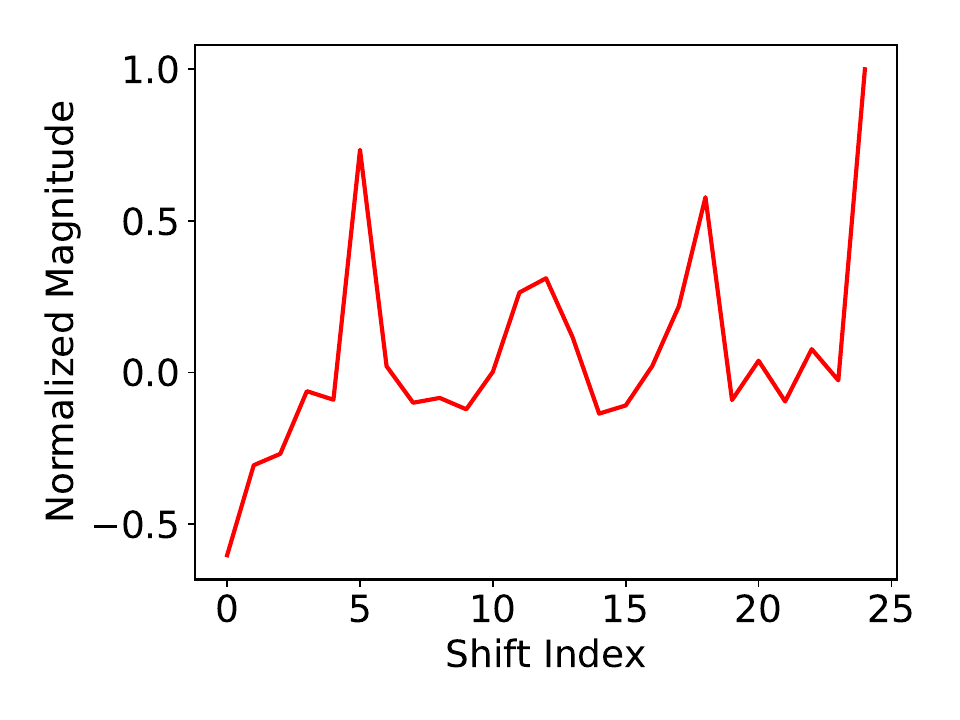}}
	\subfigure[WEARec,WFE,2 layer]{
		\includegraphics[width=0.225\textwidth]{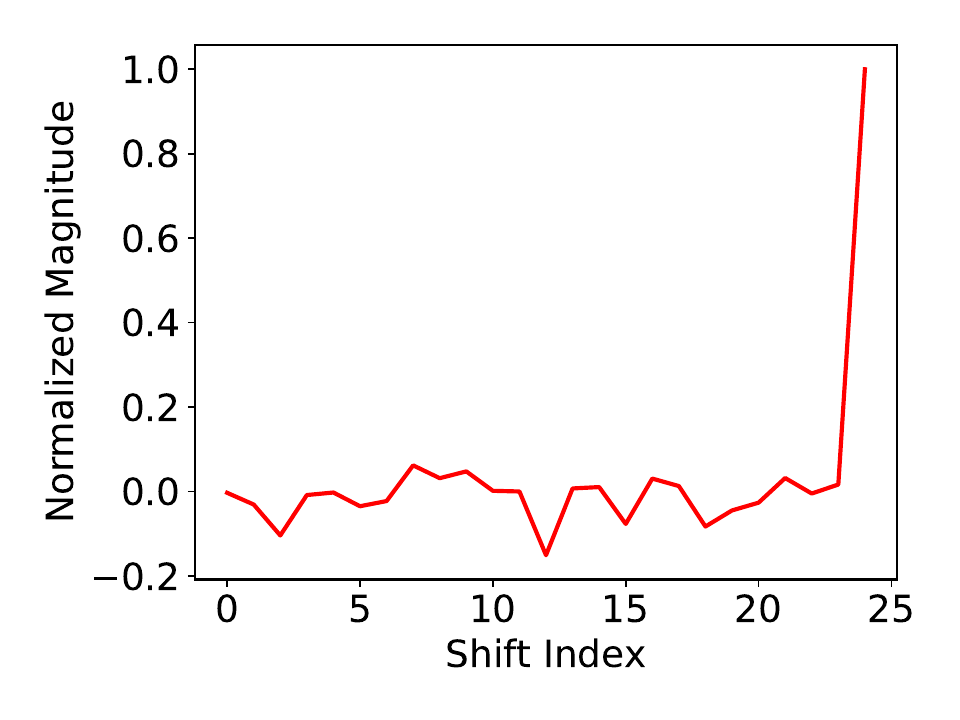}}
	\caption{Visualization of learned $\bold{T}$ in Beauty}
    \label{FigT}
\end{figure}
\begin{figure}[t]
	\centering  
	\subfigbottomskip=2pt 
	\subfigcapskip=-5pt 
	\subfigure[Sports]{
		\includegraphics[width=0.225\textwidth]{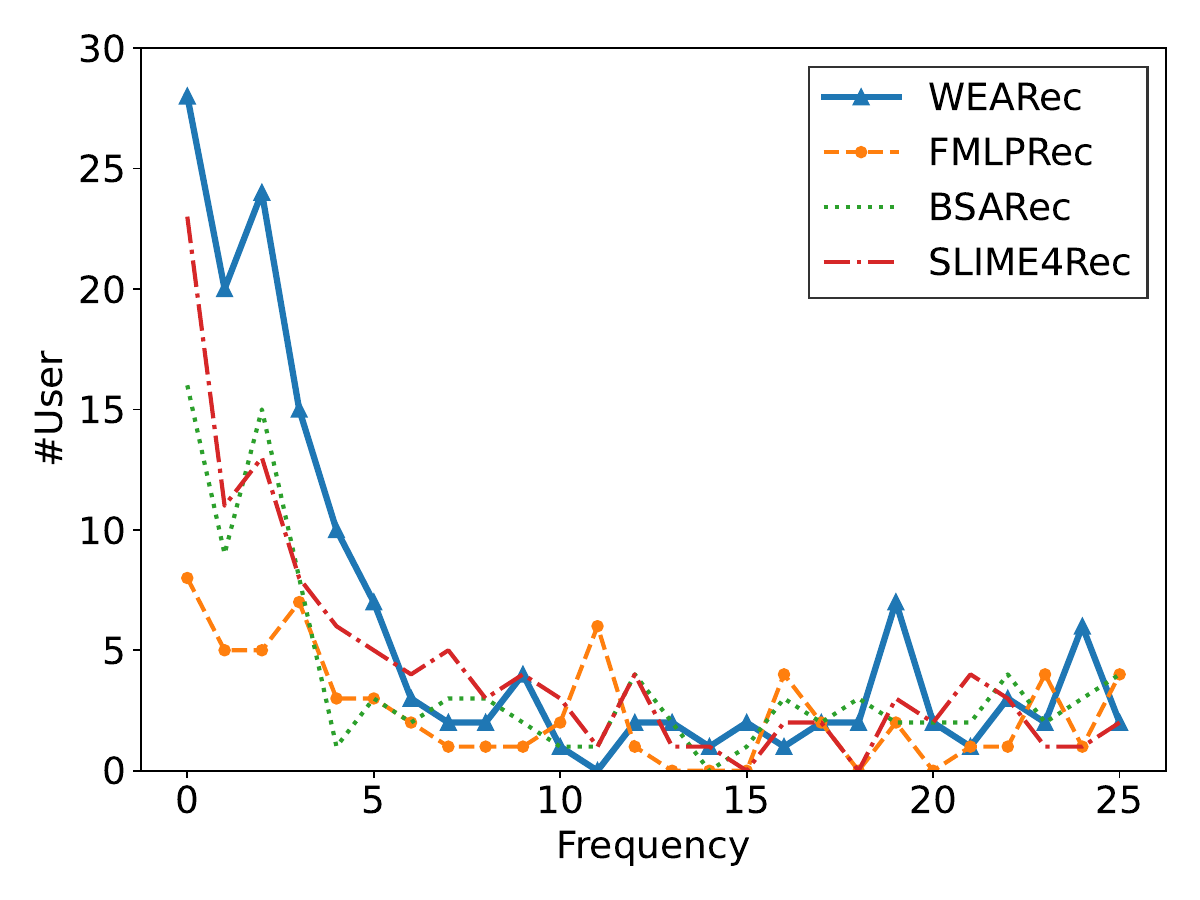}}
	\subfigure[Beauty]{
		\includegraphics[width=0.225\textwidth]{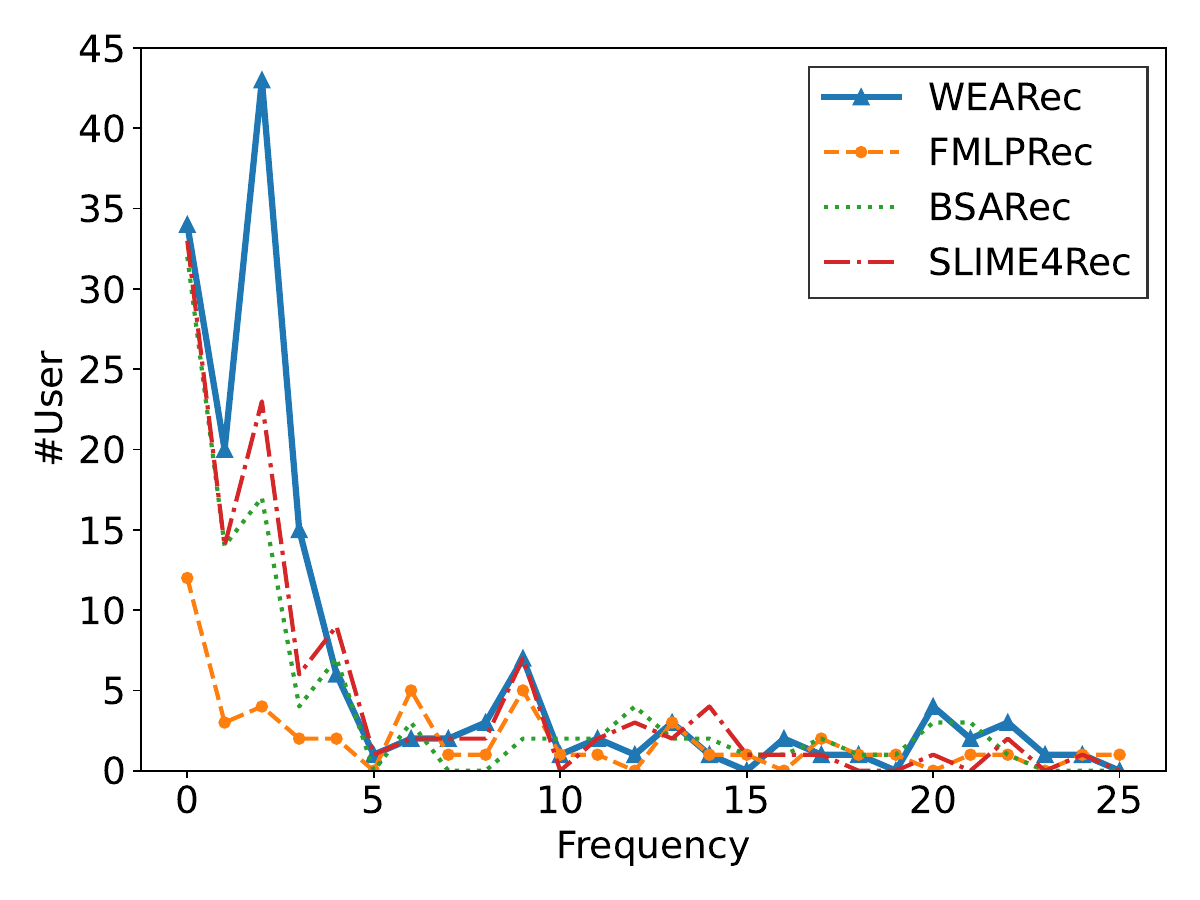}}
	\caption{Case Study in Sports and Beauty.}
    \label{FigCase}
\end{figure}

In Figure \ref{FigT}, we visualize the learned wavelet feature enhancer $\bold{T}$ in Equation 19 at each layer on the Beauty dataset. We can observe that the first-layer wavelet feature enhancement module learns multiple time points of non-stationary signals and assigns them higher weights. Moreover, for non-stationary signal points that might constitute noise information, negative weights are assigned to suppress them. In the second layer, the nearest non-stationary signal points are selected for enhancement. This indicates that the wavelet feature enhancer can capture the time points of non-stationary signals and adaptively adjust their weights.

\subsection{C.2 Case Study}
To evaluate whether the dynamic filter and wavelet enhancer can better capture a wider range of frequency domain features, we visualized the number of users correctly captured by different models, building upon the experiment shown in the paper's introduction. This further explains the effectiveness of integrating these two modules. The results are presented in Figure \ref{FigCase}. Based on our observations from Figure, we can draw the following conclusions: 
Firstly, FMLPRec captured the fewest users across multiple frequency bands. This is because its use of static filters prevents it from capturing diverse user behavior patterns. BSARec and SLIME4Rec achieved better results because they either utilize inductive bias to enhance self-attention, or use frequency ramp structures and contrastive learning to enhance user embedding representations. Finally, using dynamic filters combined with wavelet enhancers yielded even better results. Benefiting from the dynamic filter's global information capture capability and the wavelet enhancer's local detail extraction, the model achieved the best performance in low-frequency regions and certain high-frequency regions.

\typeout{get arXiv to do 4 passes: Label(s) may have changed. Rerun}

\end{document}